\documentclass[amsmath,amssymb,aps,prl,reprint,groupedaddress]{revtex4-1}

\usepackage{amsmath}
\usepackage{xcolor}
\usepackage{amsfonts,amssymb}
\usepackage{graphicx}
\usepackage{hyperref}

\begin{document}

\title{An $\ell_0\ell_2$-norm regularized regression model for construction of robust cluster expansions in multicomponent systems}


\author{Peichen Zhong}
\affiliation{Department of Materials Science and Engineering, University of California, Berkeley, California 94720, United States}
\affiliation{Materials Sciences Division, Lawrence Berkeley National Laboratory, California 94720, United States}

\author{Tina Chen}
\affiliation{Department of Materials Science and Engineering, University of California, Berkeley, California 94720, United States}
\affiliation{Materials Sciences Division, Lawrence Berkeley National Laboratory, California 94720, United States}

\author{Luis Barroso-Luque}
\affiliation{Department of Materials Science and Engineering, University of California, Berkeley, California 94720, United States}
\affiliation{Materials Sciences Division, Lawrence Berkeley National Laboratory, California 94720, United States}

\author{Fengyu Xie}
\affiliation{Department of Materials Science and Engineering, University of California, Berkeley, California 94720, United States}
\affiliation{Materials Sciences Division, Lawrence Berkeley National Laboratory, California 94720, United States}

\author{Gerbrand Ceder}
\email[]{gceder@berkeley.edu}
\affiliation{Department of Materials Science and Engineering, University of California, Berkeley, California 94720, United States}
\affiliation{Materials Sciences Division, Lawrence Berkeley National Laboratory, California 94720, United States}

\date{\today}

\begin{abstract}
We introduce $\ell_0\ell_2$-norm regularization and hierarchy constraints into linear regression for the construction of cluster expansions to describe configurational disorder in materials. The approach  is implemented through mixed integer quadratic programming (MIQP). The $\ell_2$-norm regularization is used to suppress intrinsic data noise, while the $\ell_0$-norm is used to penalize the number of non-zero elements in the solution. The hierarchy relation between  clusters imposes relevant physics and is naturally included by the MIQP paradigm. As such, sparseness and cluster hierarchy can be well optimized to obtain a robust, converged set of effective cluster interactions with improved physical meaning. We demonstrate the effectiveness of $\ell_0\ell_2$-norm regularization in two high-component disordered rocksalt cathode material systems, where we compare the cross-validation, convergence speed, and the reproduction of phase diagrams, voltage profiles, and Li-occupancy energies with those of the conventional $\ell_1$-norm regularized cluster expansion models.

\end{abstract}


\maketitle
\section{Introduction}
First-principles density functional theory (DFT) calculations have been demonstrated as a reliable tool in computational materials science. Despite the increase in computing power and accuracy of DFT methods, the scaling with the number of atoms $\sim O(n^3)$ intrinsically prohibits large-scale calculations (over $10^3$ atoms) or sampling of a high-dimensional occupancy space (millions of structures) \cite{jain2013MP}. This is particularly relevant in systems with configurational degrees of freedom that need to be sampled at non-zero temperature to equilibrate states of partial order, and their associated entropy and free energy. The state of configurational order determines many materials properties, especially in systems composed of many species (high-entropy systems). It has also recently been shown to be relevant for mechanical properties in metallic alloy systems \cite{zhang2020SRO_nature,yin2021atomistic} and the energy density of complex electrode materials for energy storage application \cite{Ji2019,Bin2020AEM_SRO_perco,Lun2020highEntropy,Zhong2020Mg}.

The cluster expansion (CE) method has been well developed to describe such configurational energetics for metallic alloys \cite{Sanchez1984, Sanchez2019}, as well as for ionic systems \cite{Tepesch1995, Seko2009selection}. The CE method expands any property (e.g., formation energy, volume) in terms of the distribution of atoms on a set of predefined sites. When the quantity being expanded is the energy, the expansion coefficients are referred to as Effective Cluster Interactions (ECI).  For example, in a multicomponent system, the energy is expanded as 
\begin{equation}
	E(\boldsymbol{\boldsymbol{\sigma}}) = \sum_{\boldsymbol{\beta}} m_{\boldsymbol{\beta}} J_{\boldsymbol{\beta}}\left\langle \Phi_{\boldsymbol{\alpha} \in \boldsymbol{\beta}} \right\rangle_{ \boldsymbol{\beta}},~ \Phi_{\boldsymbol{\alpha}} = \prod_{i=1}^N \phi_{\alpha_i}(\sigma_i).
	\label{eq:ce_definition}
 \end{equation}
A configuration $\boldsymbol{\sigma}$ represents a specific  occupancy on all the sites of the system, where $\sigma_i$ describes which species sits on the $i$-th site of the structure. The site basis function $\phi_{\alpha_i}(\sigma_i)$ transforms the occupancy variable $\sigma_i$ into a scalar value. There are typically as many (non-constant) cluster basis functions as possible occupancies on a site minus one.  The cluster basis function label  $\boldsymbol{\alpha} = (\alpha_1, \alpha_2, \alpha_3, \ldots)$ indicates a group of sites, each with a specific basis function on it, where each entry $\alpha_i$ labels the corresponding site basis function $\phi_{\alpha_i}$. Thus, the cluster basis function $\Phi_{\boldsymbol{\alpha}} = \prod_{i=1}^N \phi_{\alpha_i}(\sigma_i)$ can be obtained by taking the product of site basis functions. 

For example, the cation sublattice of a LiMnO$_2$ rocksalt oxide is a binary system, where Li and Mn share the octahedral interstitial of the FCC anion framework.  In such a system, Li can be encoded by $\sigma^{\mathrm{Li}} = 0$ and Mn by $\sigma^{\mathrm{Mn}} = 1$. The parameter $\alpha_i$ takes a value from $[0, 1, ... M-1]$, where $M$ is the number of allowed species defined on the sublattice (e.g., $M=2$ for Li-Mn). While many forms of site basis function can be used\cite{Sanchez1984,Zhang2016indicator,barrosoluque2021sparse}, a sinusoid (orthogonal) basis function is applied here to transform the occupancy variable ($\sigma^{\mathrm{Li}},\sigma^{\mathrm{Mn}}$) into a value \cite{VandeWalle2009}, where
\begin{equation}
\begin{split}
    \phi_j(\sigma_i) = \begin{cases} 
                          1 & \text{if } j=0\\
                          -\cos\left(\frac{\pi(j+1)\sigma_i}{M}\right) &\text{if } j\text{ is odd} \\
                          -\sin\left(\frac{\pi j\sigma_i}{M}\right) &\text{if } j\text{ is even}
                      \end{cases}.
\end{split}\label{eq:sinusoid}
\end{equation}
The $j$ indicates $\alpha_i$ in Eq.\eqref{eq:ce_definition} and  can take a value of 0 or 1. Thus, we have $\phi_{j=0} \equiv 1$, $\phi_{j=1}(\sigma^{\mathrm{Li}}=0)=-1$, and  $\phi_{j=1}(\sigma^{\mathrm{Mn}}=1)=1$. This situation corresponds to the spin variables used in a generalized Ising model \cite{ceder1993Ising,huang2016MAXSAT}. For systems with species number $M>2$, the basis functions take values beyond those of spin variables $\left\{-1,1\right\}$ typically used in binary CE. Some examples of other types of site-basis functions also developed for the CE method are the Chebyshev polynomials \cite{Sanchez1984} and the indicator function (point delta function) \cite{Zhang2016indicator}. 

\begin{figure*}[tb]
\centering
\includegraphics[width=\linewidth]{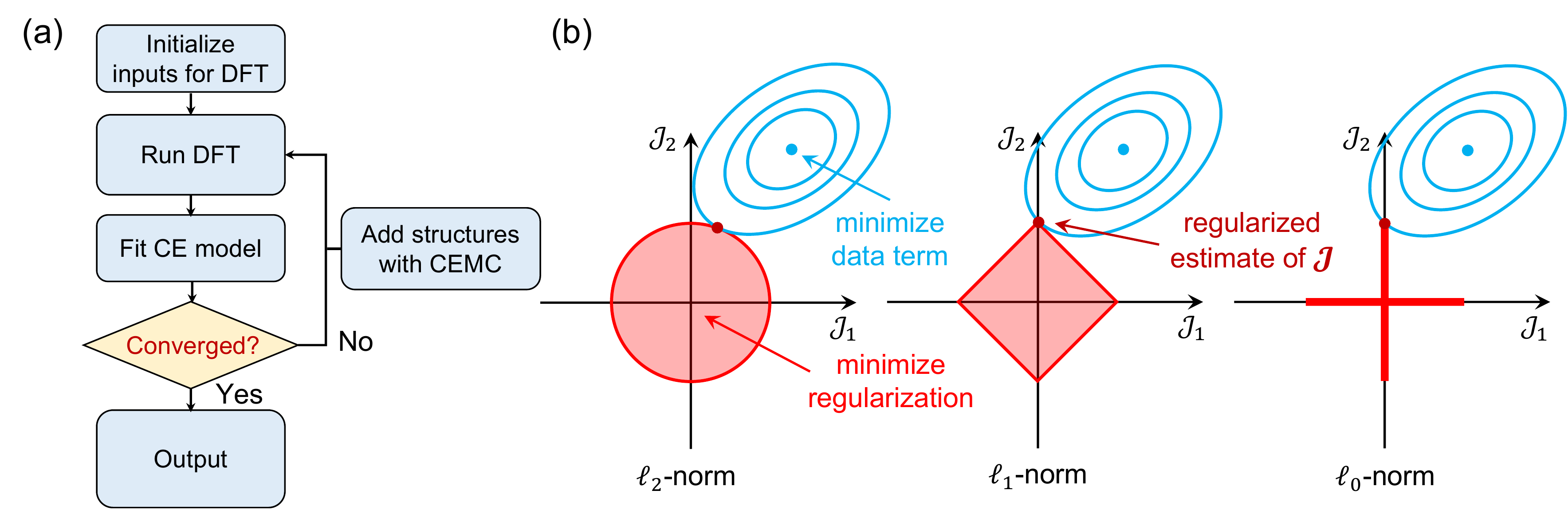}
\caption{(a) The general flowchart of constructing a CE model, including initialization of input structures, DFT calculations, fitting and convergence check, and cluster expansion Monte Carlo (CEMC) for sampling. (b) An illustration of $\ell_2$, $\ell_1$ and $\ell_0$-norm regularization in a two-parameter space $\boldsymbol{J}$ = ($J_1$, $J_2$) . The blue circles represent the contours of the data term $||\boldsymbol{E}_{\mathrm{DFT},S} -\boldsymbol{\Pi}_S\boldsymbol{J}||^2_2$ in cost function. The red regions represent the constraints of parameters (e.g. $J_1^2 +J_2^2 \leq s$ for $\ell_2$-norm, $|J_1| +|J_2| \leq s$ for $\ell_1$-norm.) The dark red point is the intersection of data term and regularization of parameters, which jointly determines the estimation of $\boldsymbol{J}$.}
\label{fig:CE_flow} 
\end{figure*}

In Eq.\eqref{eq:ce_definition}, the correlation function $\left\langle \Phi_{\boldsymbol{\alpha}} \right\rangle_{\boldsymbol{\beta}}$ is calculated by
\begin{equation}
   \langle\Phi_{\boldsymbol{\alpha}}(\boldsymbol{\sigma})\rangle_{\boldsymbol{\beta}} = \frac{1}{N_{\boldsymbol{\sigma}}m_{\boldsymbol{\beta}}}\sum_{\boldsymbol{\alpha}\in\boldsymbol{\beta}}\Phi_{\boldsymbol{\alpha}}(\boldsymbol{\sigma}),
\end{equation}
where $\boldsymbol{\beta}$ is an orbit representing all symmetrically equivalent cluster basis functions $\boldsymbol{\alpha}$, and $m_{\boldsymbol{\beta}}$ is the corresponding multiplicity. $N_{\boldsymbol{\sigma}}$ is the size of the supercell of configuration $\boldsymbol{\sigma}$; thus, the correlation function is well normalized with respect to the primitive cell. $J_{\boldsymbol{\alpha}}$ is the effective cluster interaction (ECI). We refer readers to Ref.\cite{de1979configurational, Sanchez1984, VandeWalle2009,aangqvist2019icet} for a more extended description of the CE. From Eq.\eqref{eq:ce_definition}, the CE energy is linearly dependent on the ECIs $\boldsymbol{J}$ when the configuration $\boldsymbol{\sigma}$ is fixed. We can thus write 
\begin{equation}
    E_{\mathrm{CE}}(\boldsymbol{\boldsymbol{\sigma}}) = \boldsymbol{\Pi}(\boldsymbol{\sigma})\cdot\boldsymbol{J},
    \label{eq:ce_energy}
\end{equation}
where $\boldsymbol{\Pi}(\boldsymbol{\sigma})$ is a row vector of correlation functions and $\boldsymbol{J}$ is the column vector of ECIs.

Fig.\ref{fig:CE_flow}(a) presents a brief illustration of how to iteratively construct a CE Hamiltonian. In practice, the CE model is initially fitted on a small set of DFT calculations. Then, a simple CE is fitted that can be used in a Monte Carlo simulation to sample new structures. DFT calculations will be applied to a sample of the MC-obtained structures, and a new CE will be fitted. This procedure will be performed iteratively until the model is converged (i.e., the cross-validation error remains low and stable, the model reproduces DFT ground states well, etc.) \cite{Huang2017GSpreserve}. In such a process, it is always desirable to achieve fewer training iterations, as DFT calculations are costly in terms of CPU time. On the other hand, fewer structures may also result in a worse fitting due to insufficient sampling of the configuration space. 

Obtaining reliable ECIs $\boldsymbol{J}$ from the DFT energy of a set of configurations is the central problem of CE fitting. Given a set of input occupancy configurations $S$, the set of correlation vectors forms a feature matrix $\boldsymbol{\Pi}_S = [\boldsymbol{\Pi}_1, \boldsymbol{\Pi}_2, ...]$, and the corresponding DFT energies are used to construct the target vector $\boldsymbol{E}_{\mathrm{DFT},S}$. Determining the ECIs is an inverse problem of Eq.\eqref{eq:ce_energy}, also called linear regression. Generally, the problem can be solved by minimizing the cost function
\begin{equation}
    \min_{\boldsymbol{J}} ~ ||\boldsymbol{E}_{\mathrm{DFT},S} -\boldsymbol{\Pi}_S\boldsymbol{J}||^2_2 + \mu||\boldsymbol{J}||_p ~,~||\boldsymbol{x}||_p = \left( \sum_i |x_i|^p\right)^{\frac{1}{p}},
    \label{eq:obj_function}
\end{equation}
where the $p$-norm of $\boldsymbol{J}$ is added to regularize the fit and suppress over-fitting, and $\mu$ controls the degree of regularization. Fig.\ref{fig:CE_flow}(b) shows the comparison of $\ell_2$, $\ell_1$ and $\ell_0$-norm regularization in a two-parameter space $\boldsymbol{J}$ = ($J_1$, $J_2$).  The blue circles are the contours of the data term error $||\boldsymbol{E}_{\mathrm{DFT},S} -\boldsymbol{\Pi}_S\boldsymbol{J}||^2_2$. The red regions represent the regularization constraints on the parameters ($||\boldsymbol{J}||_p \leq s$), which can be transformed to a Lagrangian form $\mu ||\boldsymbol{J}||_p$ as shown in Eq.\eqref{eq:obj_function}. The dark red point is the regularized estimation of $\boldsymbol{J}$, which is the intersection between the data error term and the regularization term. The $\ell_1$-norm tends to generate sparser solutions compared with the $\ell_2$-norm, because the intersection is likely to be located on the axis. The $\ell_0$-norm counts the non-zero elements of $\boldsymbol{J}$, where the intersection is exactly located on the axis and thus the $\ell_0$-norm imposes an exact sparsity constraint on $\boldsymbol{J}$. 

Conventionally, $\ell_2$-norm (\textit{ridge regression}, $p=2$) regularization can be applied when the problem is over-determined (i.e., the number of training structures is larger than the dimension of $\boldsymbol{J}$). The $\ell_2$-norm regularized regression reduces the over-fitting caused by intrinsic noise in the training data. This can be achieved solely by introducing the $\ell_2$ regularization function and additionally using the mixed-basis expansion \cite{PhysRevB.46.12587_Laks_1992,PhysRevB.70.155108_Zunger_2004,Mueller2009}. Bayesian approaches have also been successfully applied to estimate the ECIs with a prior distribution in several binary systems \cite{Nelson2013_BayesianCS, PhysRevB.81.012104_Axel_sparse_precise,Mueller2009}. However, the number of ECIs increases combinatorially with the number of species, scaling approximately as $\prod_{k} (M_k-1)^{n_k}$, where $M_k$ is the number of species on the $k$-th sublattice, and $n_k$ is the number of cluster sites in the same sublattice $k$. The explosion in the number of basis functions when many species can occupy a site makes it difficult to predefine which cluster basis functions contribute to the expansion for high dimensional multi-component systems (i.e., which cluster basis function has a non-zero element in the solution of $\boldsymbol{J}$). Therefore, a sparse solver for ECIs selection is required.

Rigorously, the exact sparse solution of Eq.\eqref{eq:obj_function} is obtained with $\ell_0$-norm regularization of $\boldsymbol{J}$. However, it is hard to compute the $||\boldsymbol{J}||_0$ in the cost function as it is an NP-hard problem. In the compressive sensing paradigm, the $\ell_0$-norm  can be transformed to an $\ell_1$-norm when the feature matrix $\boldsymbol{\Pi}_S$ satisfies the restricted isometry property (RIP) condition \cite{Candes2008}. To satisfy the RIP condition in CE, \citet{Nelson2013} proposed to generate a training set in which each row is an identical independent distributed (i.i.d.) random vector. However, in practical cases, the configurations in the training set $S$ are correlated, because structures are not randomly sampled, but are mostly part of an ensemble of configurations with low energy. Such correlations fail to satisfy the i.i.d. condition. Moreover, generating structures from a specific correlation vector is also an NP-hard problem. Though the strict compressive sensing cluster expansion is not easy to construct in practice, the $\ell_1$-norm (\textit{lasso}, $p=1$) regularization is widely used as feature selection, which has shown success in various alloy and ionic systems \cite{Nelson2013, Seko_sparseL1, Leong2019grouplasso, barrosoluque2021sparse}. 

In this paper, we propose an $\ell_0\ell_2$-norm regularization approach that incorporates hierarchy constraints to generate more robust and predictive CE models. First, we introduce the $\ell_0\ell_2$ penalty term and hierarchy constraints in the paradigm of mixed-integer quadratic programming (MIQP). Second, we compare the sparseness and convergence rate of ECIs with those of the conventional $\ell_1$ method in the Li-Mn-V-Ti-O-F disordered rocksalt system. Finally, we demonstrate that an $\ell_0\ell_2$-regularized CE better reproduces the correct physical interactions by comparing with $\ell_1$-CE in terms of computed phase diagrams, voltage profiles, and related physical quantities in the Li-Mn-Ti-O system.

\section{Methods}
\subsection{The $\ell_0$-norm regularization }
In Eq.\eqref{eq:obj_function}, $p=0$ manifests itself as a pseudo-norm that counts the number of non-zero elements of $\boldsymbol{J}$:
\begin{equation}
    ||\boldsymbol{J}||_0 = \sum_i \mathrm{Ind}(J_i), ~\mathrm{Ind}(J_i) = \left\{ 
    \begin{aligned}
        &0, J_i = 0\\
        &1, J_i \neq 0
    \end{aligned}
    \right.
\end{equation}
Adding the $\ell_0$ term into the cost function directly penalizes the number of non-zero ECIs, yielding better sparseness in its solution. However, optimizing a cost function with an $\ell_0$ term is an NP-hard problem and is difficult to present in a direct way \cite{Candes2008,elad_pursuit_2010}. Previously, \citet{Huang2018} has approached the problem by rewriting $\ell_0$ optimization as a mixed-integer programming problem, such that
\begin{align}
\min ||\boldsymbol{J}||_0 \quad \Leftrightarrow \quad &\min ~\sum_{ c\in \boldsymbol{C}} z_{0,c} \\
\nonumber
&\begin{array}{r@{~}l@{}l@{\quad}l}
\boldsymbol{s.t.} \quad & Mz_{0,c} \geq J_c, ~\forall c\in \boldsymbol{C}\\
& Mz_{0,c} \geq -J_c,~\forall c\in \boldsymbol{C}\\
& z_{0,c} \in \{0,1\}, ~\forall c\in \boldsymbol{C}
\end{array} 
\end{align}
where $M$ is a sufficiently large number (larger than the maximum possible absolute value of any ECI), and $z_{0,c}$ is a slack variable (binary integer) indicating whether the ECI of orbit $c$ is zero or not. $J_c$ is constrained to $0$ when the slack variable $z_{0,c} = 0$ (inactive) and to $[-M,M]$ when $z_{0,c} = 1$ (active). (For a rigorous mathematical background, refer to Ref.\cite{Huang2018}.) In practice, it is shown that one can at least obtain a sparseness-improved near-optimal solution within a reasonable CPU time cutoff. In our benchmark tests of the Li-Mn-V-Ti-O-F and Li-Mn-Ti-O systems, the optimizations of ECIs were completed within 600s using the \textit{gurobi} package \cite{gurobi}.

\subsection{Hierarchy constraints}
In a CE, clusters are usually enumerated in an iterative, low-to-high order (i.e., from singlets to pairs, triplets, quadruplets, and so on). Practically, the CE is truncated to a maximum of $n$ (e.g., quadruplet clusters with $n=4$ are a typical limit), ignoring the higher-order interactions to control the model complexity. To differentiate the cluster orbits by different significance, we take one of the basic assumptions of CE that $n$-body cluster interactions become less important to the configurational energy (or other scalar properties) as $n$ becomes larger. This assumption means that the majority of the fitted property can be described by the lower-order interactions and that the higher-order interactions serve as the fine-tuning part in the fitting. 

Such a physically inspired concept can be introduced in the form of hierarchy constraints, as has been done successfully in some previous studies \cite{PhysRevB.71.212201_hierarchy, PhysRevLett.92.255702_hierarchy,VandeWalle2002Automate}. The hierarchy constraint manifests itself as $J_b \neq 0$ if and only if $J_a \neq 0$ ($a\subset b$), where $a$ and $b$ are a lower- and higher-order cluster function orbit, respectively, and $b$ contains all the site bases of $a$ as a subset. In the MIQP representation, the hierarchy relationship can be easily expressed as a constraint between slack variables:
\begin{equation}
    z_{0,b} \leq z_{0,a}, ~ a\subset b.
\label{Eq:constrain}
\end{equation}
This treatment was first proposed by \citet{Huang2018}, where it was used in the $\ell_0\ell_1$-norm regularization paradigm. 

\subsection{The $\ell_2$-norm regularization}

We propose that combining $\ell_2$-norm and $\ell_0$-norm regularization can impose true hierarchy constraints unlike the $\ell_0\ell_1$-norm. It is to be noted that the inequality between slack variables does not necessarily impose the hierarchy relation ($J_b\neq 0$, iff $J_a\neq0$). This is because the hierarchy constraints are defined on the magnitude of ECIs $J_a$ and $J_b$, while the slack variables  $z_{0,b}, z_{0,a}$ are intermediate to represent the presence or exclusion of the variables.

\begin{figure}[htb]
\includegraphics[width=0.9\linewidth]{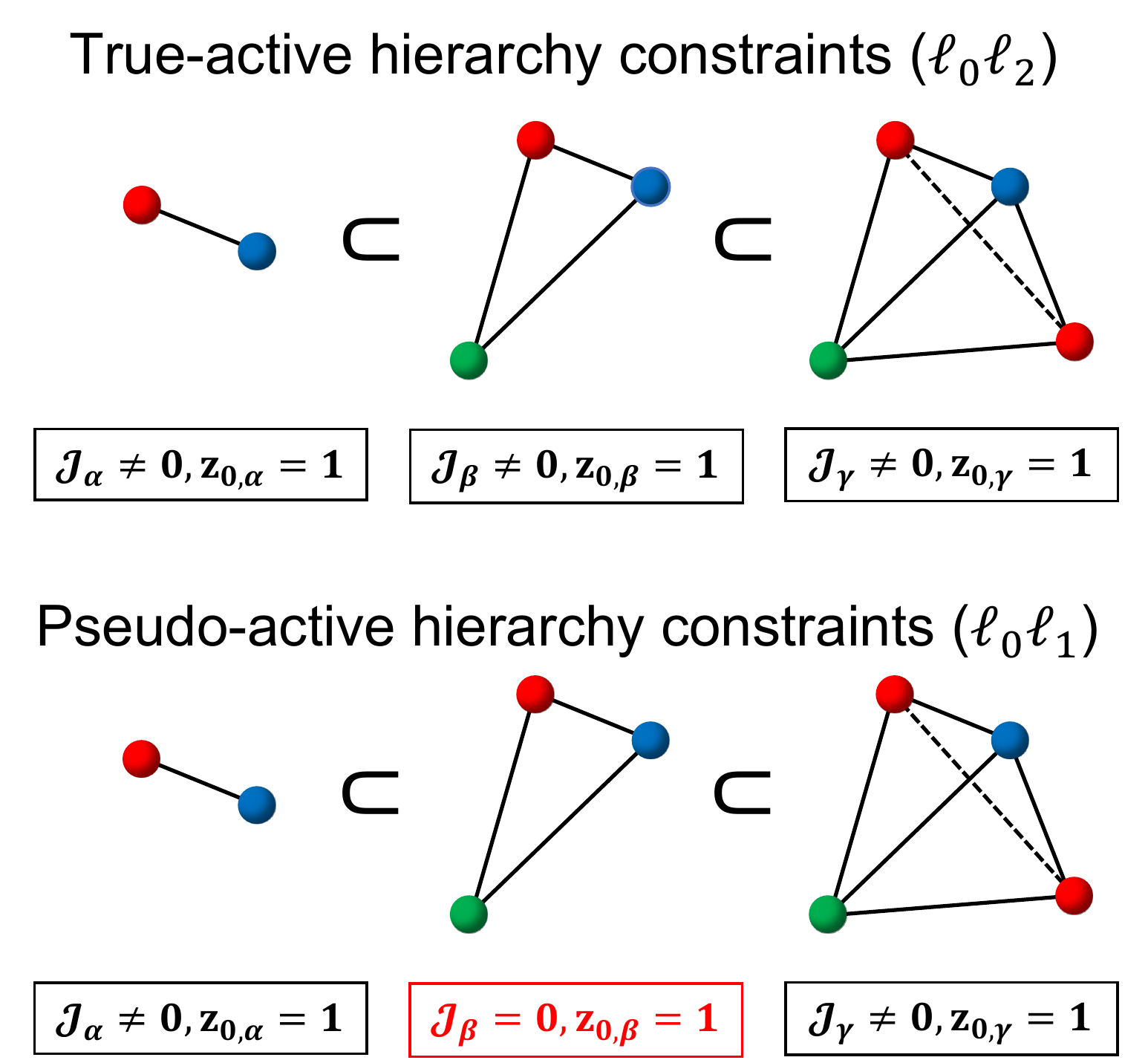}
\caption{Illustration of hierarchy relations ($\alpha\subset \beta \subset \gamma$) between pair, triplet, and quadruplet orbit. The different colors on the cluster sites represent the decorating species for a given site-basis function. The equation in red shows a pseudo-active hierarchy constraint that may appear in $\ell_1$ and its derivative methods.}
\label{fig:CE_hierarchy} 
\end{figure}

When implementing the hierarchy constraints in $\ell_0\ell_1$-norm regularization, pseudo-active behavior can manifest itself when a $J = 0$, but its slack variable $z_{0} = 1$ within the MIQP paradigm. $J$ can be regularized to zero, which is still a valid solution between $[-M,M]$, even with $ z_{0} = 1$. This is caused by the fact that the $\ell_1$-norm has feature-selection properties that intrinsically produce a sparse solution \cite{tibshirani1996regression}.  This pseudo-activeness can introduce excessive sparseness to the solution and break the hierarchy constraints. Fig.\ref{fig:CE_hierarchy} presents an example of pseudo-activeness in $\ell_0\ell_1$-norm regularization. The excessive sparseness is introduced to the orbit $\beta$ with $J_\beta = 0$, while all orbits $\alpha, \beta, \gamma$  has active slack variables $z_{0}=1$. The higher-order orbit $\gamma$ is erroneously activated while $J_\beta = 0$. To avoid such a situation and ensure proper function with $\ell_0$ under hierarchy constraints, a norm with no feature-selection properties is required. The $\ell_2$-norm is a natural choice.

With the introduction of the $\ell_0\ell_2$-norm and hierarchy constraints, the final ECI optimization problem can be written as
\begin{align}
&\min_{\boldsymbol{J}} ~ \boldsymbol{J}^T\boldsymbol{\Pi}_S^T\boldsymbol{\Pi}_S\boldsymbol{J}^T - 2\boldsymbol{E}_{\mathrm{DFT}}^T\boldsymbol{\Pi}_{S}\boldsymbol{J} + \mu_0\sum_{ c\in \boldsymbol{C}} z_{0,c} +\mu_2||\boldsymbol{J}||_2^2 \label{eq:corrL0L2}\\
\nonumber
&\begin{array}{r@{~}l@{}l@{\quad}l}
\boldsymbol{s.t.} \quad Mz_{0,c} &\geq J_c, ~\forall c\in \boldsymbol{C}\\
Mz_{0,c} &\geq -J_c,~\forall c\in \boldsymbol{C}\\
z_{0,b} &\leq z_{0,a},  ~\forall a \subset b, ~ \{a,b\} \in \boldsymbol{C}\\
z_{0,c} &\in \{0,1\}, ~\forall c\in \boldsymbol{C},
\end{array}
\end{align}
where $||\boldsymbol{J}||_2^2 = \boldsymbol{J}^T\boldsymbol{J}$ penalizes the magnitude of ECIs, thus avoiding over-fitting by regularizing sampling noise while the $\ell_0$-term $\sum_{c} z_{0,c}$ optimizes the sparseness. The hierarchy constraints ensure correct containment relationships by manifesting lower-order ECIs first to reduce redundancy. The packages for our implementation are available in Ref.\footnote{\href{https://github.com/CederGroupHub/smol}{https://github.com/CederGroupHub/smol}, \href{https://github.com/CederGroupHub/sparse-lm}{https://github.com/CederGroupHub/sparse-lm}}.

\section{Results}
\begin{figure}[h!]
\centering
\includegraphics[width=\linewidth]{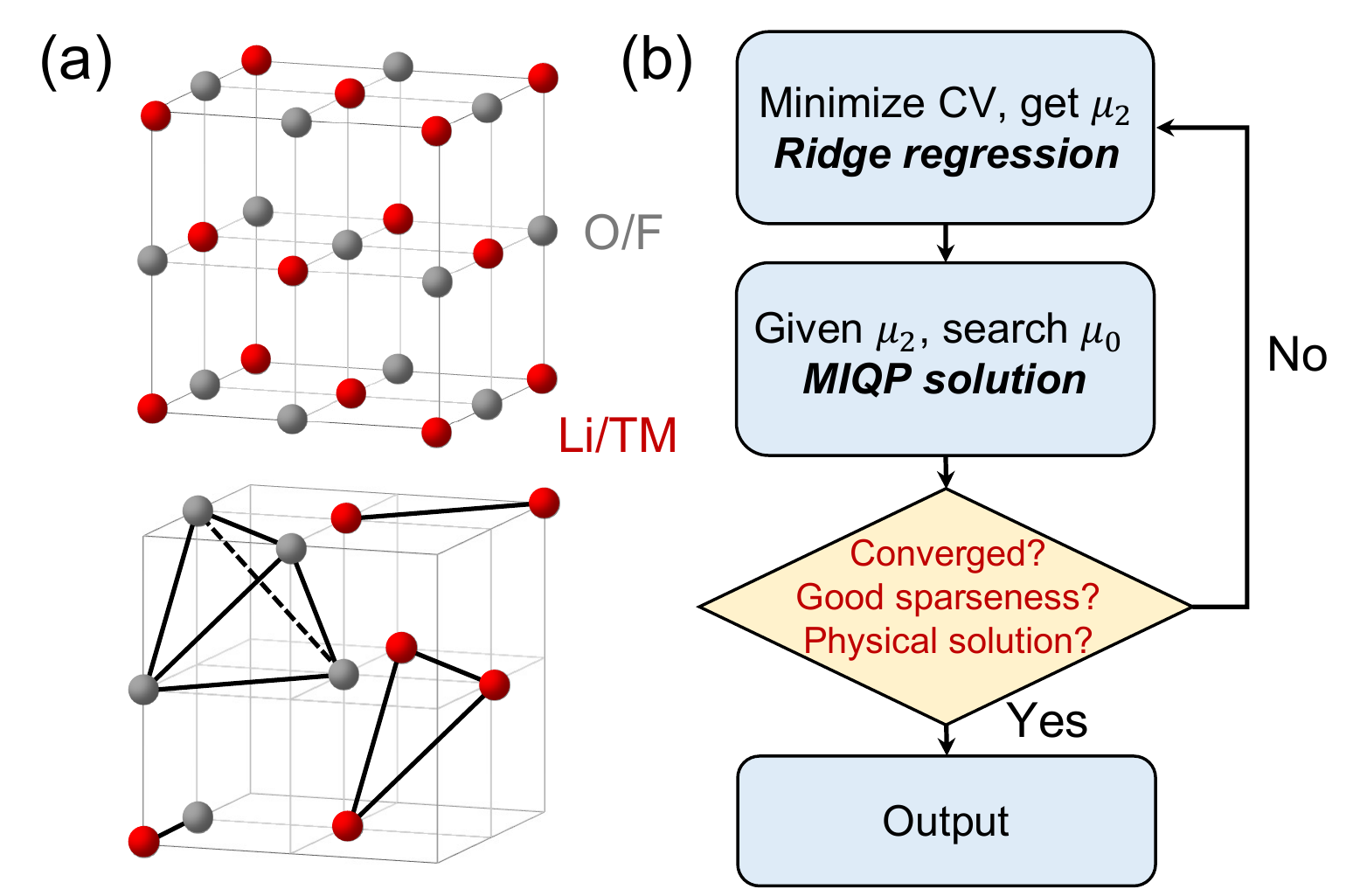}
\caption{(a) An illustration of the rocksalt lattice structure. The cation sites are labeled in red and can be occupied by Li$^+$ and transition metals (TM, including Mn$^{2+}$, V$^{3+}$, Ti$^{4+}$ in our example) in DRX. The anion sites are labeled in gray and can be occupied by O$^{2-}$ and F$^{-}$. The lower panel gives some examples of $n$-body ($n=2,3,4$) clusters used in the CE model, including intra and inter-sublattice interactions. (b) The procedure to obtain an $\ell_0\ell_2$-norm regularized solution, including finding the $\mu_2$ by minimizing the CV error in ridge regression, sparseness engineering with $\ell_0$ using MIQP, and terminating if the solution is converged with good sparseness, as well as good-reproduction-relevant physical properties. }
\label{fig:structure_solver} 
\end{figure}
As mentioned above, for systems with many species, the number of basis functions grows rapidly. An example of such systems are the Li-excess disordered rocksalts (DRX), which are multicomponent systems that can be synthesized with a wide variety of elements \cite{Clement2020}. Recently, high-entropy DRX materials have been synthesized with up to 12 metal species \cite{Lun2020highEntropy}. In addition, their configurational short-range order is critical to their transport properties, warranting a detailed CE approach.\cite{Huang2021NatureEnergy,Bin2020AEM_SRO_perco} Here, we provide a heuristic solution to study configurations in such high-dimensional  DRX systems by applying an $\ell_0\ell_2$-regularized CE model to fit the formation energy of DRX compounds. 

\subsection{Robustness and convergence}
The convergence of the CE when the $\ell_0\ell_2$-norm and hierarchy constraints are enforced was tested on configurational disorder in the LiF--MnO--LiVO$_2$--Li$_2$TiO$_3$ composition space. The CE model contains pair interactions up to 7.1 \AA, triplets up to 4.0 \AA, and quadruplets up to 4.0 \AA\ based on a lattice parameter $a=3$ \AA\ for the primitive cell. Fig.\ref{fig:structure_solver}(a) presents the rocksalt framework of a DRX structure. The framework contains a cation sublattice (red) and anion sublattice (gray), where the cation sites can be occupied by Li and transition metals (TM, including Mn$^{2+}$, V$^{3+}$, Ti$^{4+}$ in this example) and the anion sites can be occupied by O$^{2-}$ and F$^{-}$. A species indicator where the site basis function reads $\phi_{j}(\sigma_i)=\delta_{i,j}$ was used \cite{Zhang2016indicator}. The electrostatic energy (Ewald energy $E_0/\varepsilon_r$) is also included to capture long-range electrostatic interactions ($E_0$ is the unscreened electrostatic energy and $1/\varepsilon_r$ is fitted as one of the ECI  ($1 /\varepsilon_r \geq 0$) \cite{seko2014electrostatic,Richards2018Fluorination}. In total, 162 ECIs (including the constant term $J_0$) are predefined in the CE Hamiltonian. The dimension of the feature matrix $\boldsymbol{\Pi}_{\text{DFT},S}$ is $487\times 162$. The performance of the $\ell_0\ell_2$-CE is compared with the  $\ell_1$-CE. We emphasize two major improvements in the $\ell_0\ell_2$-CE. 

\begin{figure}[htb]
\centering
\includegraphics[width=\linewidth]{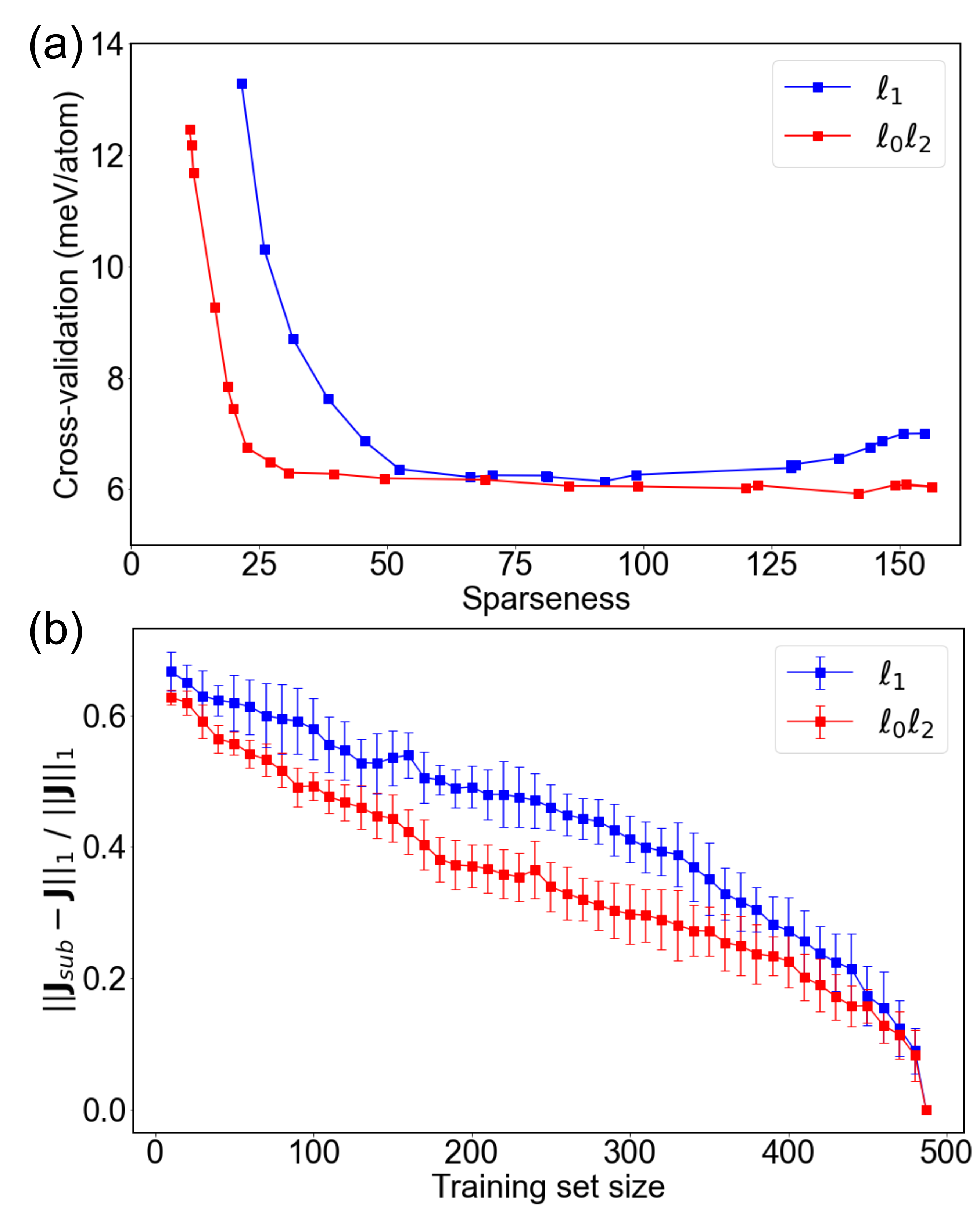}
\caption{(a) Cross-validation error (meV/atom) of the $\ell_1$-CE and the $\ell_0\ell_2$-CE. The sparseness is the number of non-zero ECIs in the fit ($||\boldsymbol{J}||_0$). The curves are generated by varying hyperparameters $\mu_0, \mu_1$, and $\mu_2$ in regularization. (b) ECIs convergence test vs training set size. $\boldsymbol{J}$ is the ECIs fitted with full training data, and $\boldsymbol{J}_{sub}$ is the ECIs fitted with a subset of corresponding size.}
\label{fig:CV_L1vsL0L2} 
\end{figure}

\subsubsection{(1) Sparseness vs. cross-validation error}
Cross-validation (CV) error vs. model complexity is a general metric used to evaluate the robustness of a CE model. The optimal trade-off between under-fitting and over-fitting can be found with a CV test, where the optimal model is fitted with the regularization hyper-parameter $\mu$ that minimizes the CV error.  In our test, a $k$-fold CV error is used, 
\begin{equation}
    \text{CV} = \sqrt{\frac{1}{k}\sum_{j=1}^k \text{MSE}_j}, ~\text{MSE} = \frac{1}{N}\sum_{i=1}^N (E_{\text{DFT}}^i-E_{\text{CE}}^i)^2,
\end{equation}
where CV is the cross-validation error averaged over $k$ splits of the validation dataset, and MSE is the mean-squared-error of each validation dataset.
Here, $N$ is the size of the validation dataset, and $k=5$ is the number of folds.  In our tests, the regularization hyperparameter $\mu$ is selected from the logarithm space between $[10^{-6},10^{-1}]$. The sparseness is defined as the number of non-zero elements of the solution ($||\boldsymbol{J}||_0$) and represents the model complexity.

The CV error versus sparseness is presented in Fig.\ref{fig:CV_L1vsL0L2}(a) for an $\ell_1$ and $\ell_0\ell_2$-norm regularized CE. For the $\ell_0\ell_2$-CE, the CV error remains low as the sparseness varies between 25 and 150 ECIs. In this regime, the $\ell_0\ell_2$-CE shows no sign of over-fitting as the CV error remains near the global minimum around 6 meV/atom. The $\ell_1$-CE shows a similar optimal CV error as that of $\ell_0\ell_2$-CE near this \emph{minimum plateau} regime from 50 to 100 in sparseness. However, as the model complexity changes, the CV error increases at both low and high sparsity, indicating that the $\ell_1$-CE is less robust against the choice of model complexity. Therefore, we conclude that the $\ell_0\ell_2$-CE can reach low CV error with a lower complexity, which is empirically believed to result in models that better reproduce physics. A more sparse CE can increase the computational speed of energy evaluations and is also less sensitive to model complexity change as compared with the $\ell_1$-CE.

\subsubsection{(2) Convergence of ECIs with a subset of training data}

The second point that we want to emphasize is that the $\ell_0\ell_2$-CE converges to its most accurate solution faster than the $\ell_1$-CE, which lowers the risk of obtaining an over-fitted result when the configuration space is under-sampled. This is an important improvement in the practical use of CE constructions. To test this hypothesis and mimic the iterative sampling process, we designed a numerical experiment based on a finished DFT dataset (with 487 structures in total). Then, we evaluated the quality of fits performed on subsets of training data of increasing size. We subsequently compared the subset-fitted ECIs $\boldsymbol{J}_{sub}$ with the full-set result. In such a comparison, the ground-truth (full set) solution is set as follows: (1) For the $\ell_1$-CE, the regularization parameter $\mu_1$ is chosen at the minimum CV error according to Fig.\ref{fig:CV_L1vsL0L2}(a). This solution has 99 non-zero ECIs when all 487 training structures are used in the fitting. (2) For the $\ell_0\ell_2$-CE, to compare the convergence rate under a similar degree of model complexity, hyperparameters are selected such that the $\ell_1$-CE and $\ell_0\ell_2$-CE have similar sparsity. The resulting $\ell_0\ell_2$-CE has 92 non-zero ECIs with all 487 training structures included according to Fig.\ref{fig:CV_L1vsL0L2}(a).  

After setting the hyperparameters for both models, we compared the normalized absolute difference $||\boldsymbol{J}_{sub} - \boldsymbol{J}||_1 / ||\boldsymbol{J}||_1$ between the $\ell_1$-CE (blue line) and  $\ell_0\ell_2$-CE (red line) in Fig.\ref{fig:CV_L1vsL0L2}(b). For each subset size, ten randomly selected  subsets with the same size were evaluated and averaged. The solid square represents the average, and the error bar represents the standard deviation resulting from different subsets. Fig.\ref{fig:CV_L1vsL0L2}(b) indicates that the $\ell_1$-CE demonstrates higher deviation from the ground-truth solution and converges more slowly to it than the $\ell_0\ell_2$-CE as the training set is increased. This result unambiguously demonstrates the robustness of $\ell_0\ell_2$-CE to work with small input data sets.

\begin{figure*}[t!]
\centering
\includegraphics[width=\linewidth]{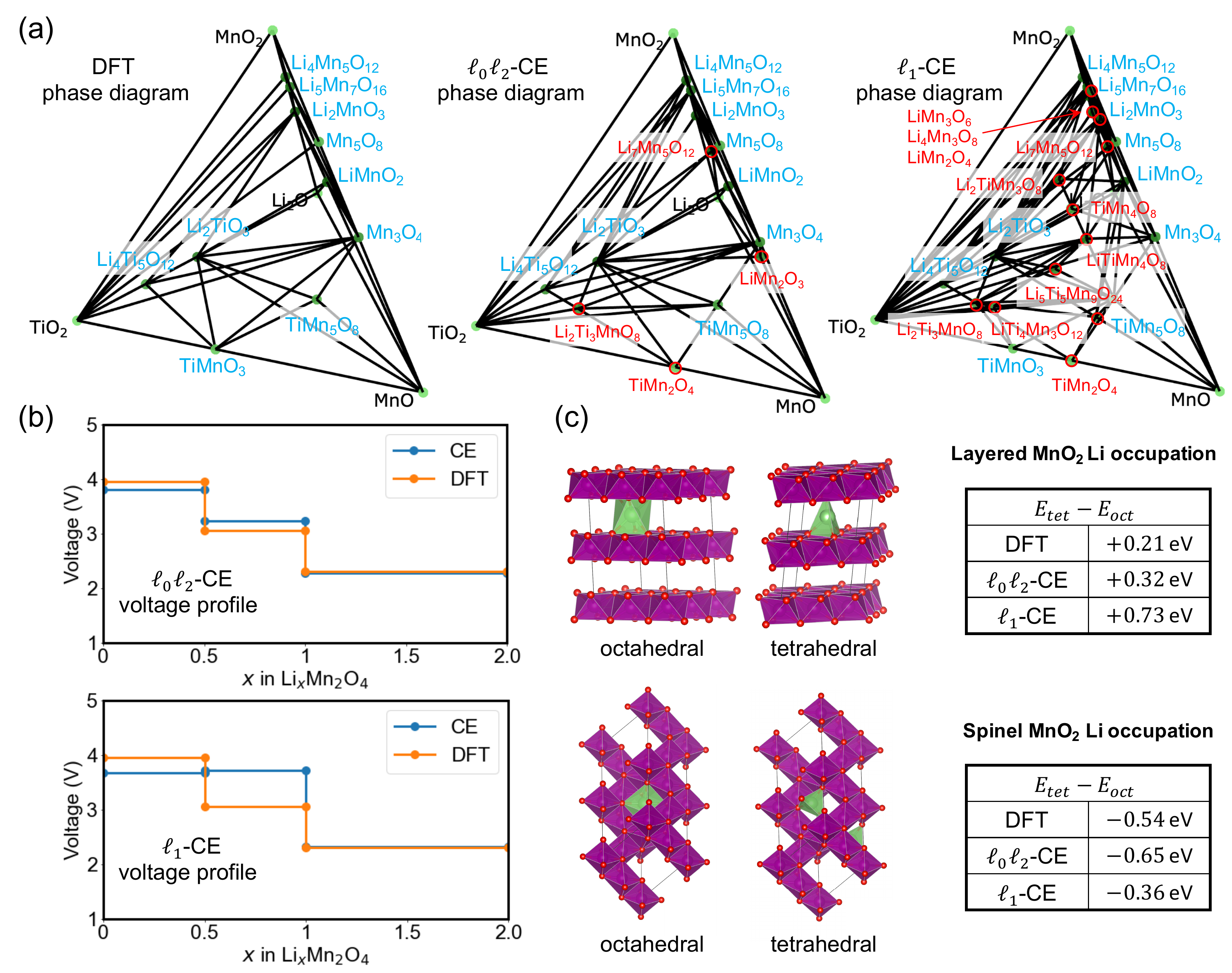}
\caption{(a) Phase diagram generated with DFT, $\ell_0\ell_2$-CE, and $\ell_1$-CE. The DFT ground states are labeled in blue text. The incorrectly predicted ground states are labeled with red circles and text. (b) The simplified spinel voltage profile (blue line) generated by $\ell_1$-CE and $\ell_0\ell_2$-CE for spinel orderings in Li$_x$Mn$_2$O$_4$ is compared with the DFT ground-truths (orange line). (c) Energy difference of Li occupation in octahedral and tetrahedral sites in layered MnO$_2$ (upper) and spinel MnO$_2$ framework (lower).}
\label{fig:tina_CE} 
\end{figure*}

\subsection{ECIs with improved physics}
From a general perspective of machine learning (ML), the predictions of energies are made by fitting statistical models on a group of data points. The statistical models can predict the absolute energy with high accuracy by minimizing the cost function, which is constructed by the difference between prediction and observation of the training data. However, in materials science, relative energy quantities are of greater significance than the absolute one (such as energy above the hull, phase diagram and the derivatives of formation energy with respect to the compositional variables). \citet{bartel2020critical} critically examined several ML models for energetics prediction, and found that while the models predict the formation energy ($\Delta H_f$) of materials well, they failed to predict the relative phase stability. Such a dilemma indicates that the prediction error (CV or RMSE) is not the only thing one should consider when constructing a statistical model for the energy. 

To demonstrate that the $\ell_0\ell_2$-CE also leads to a more physically informed solution, we studied a multicomponent system: Li-Mn-Ti-O oxide in an fcc rocksalt framework, with Li$^+$-Mn$^{2+}$-Mn$^{3+}$-Mn$^{4+}$-Ti$^{4+}$-vacancy disorder on the octahedral cation sites and Li$^+$-Mn$^{2+}$-Mn$^{3+}$-vacancy disorder on the interstitial tetrahedral sites. The Li-Mn-Ti-O composition space contains a number of battery-relevant systems \cite{Ji2019,Bin2020AEM_SRO_perco}. These battery systems are charged and discharged by adding or removing lithium (i.e., lithiation or delithiation) and a charge-compensating electron, which reduces or oxidizes a transition metal.  As a result, an important physical property to correctly model in the Li-Mn-Ti-O system is the energetics of Li in octahedral vs. tetrahedral sites. One significant battery-relevant system in which the effects of Li local environment preference are especially presented is the LiMn$_2$O$_4$ spinel. When fully lithiated to Li$_2$Mn$_2$O$_4$, Li-ions occupy octahedral sites while the Li-ions occupy tetrahedral sites for compositions Li$_x$Mn$_2$O$_4$ when $x\leq 1$.

Thus, we design two additional tests to ensure that the CE models well represent the physics of the Li octahedral vs. tetrahedral site preferences. Specifically, we compare how well the CE model reproduces: (1) energy differences between the Li in the tetrahedral and octahedral sites in layered MnO$_2$ and spinel MnO$_2$ frameworks and (2) a simplified spinel voltage profile against the DFT ground truths. The simplified spinel voltage profile includes the fully lithiated rocksalt-like Li$_2$Mn$_2$O$_4$, the spinel LiMn$_2$O$_4$, the commonly seen Li$_{0.5}$Mn$_2$O$_4$ ordering, and the fully delithiated Mn$_2$O$_4$ and is calculated by taking the average voltage between each set of adjacent orderings. The average voltage is calculated using DFT and the following equation \cite{Urban2016npj,Seo2015PRB, aydinol1997JES}:
\begin{equation}
	\Bar{V}(x_1, x_2) \approx -\frac{E_{\mathrm{Li}_{x_1}\mathrm{Mn}_2\mathrm{O}_4} - E_{\mathrm{Li}_{x_2}\mathrm{Mn}_2\mathrm{O}_4} - (x_1 - x_2)E_{\mathrm{Li}}}{F(x_1 - x_2)},
	\label{eq:avgvoltage}
 \end{equation}
where $x_1$ and $x_2$ are adjacent Li contents with $x_1 > x_2$, $E_{\mathrm{Li}}$ is the DFT energy of bcc Li metal, and $F$ is the Faraday constant.

The CE was generated with pair interactions up to 7.1 \AA, triplet interactions up to 4.0 \AA, and quadruplet interactions up to 3.0 \AA\ based on a primitive cell of the rocksalt structure with lattice parameter $a=3$ \AA. A sinusoid site basis was used (Eq.\eqref{eq:sinusoid}). 
In total, 1475 ECIs (including the constant term $J_0$) were predefined in the CE Hamiltonian. The dimension of the feature matrix is $1137\times 1475$. Because of the high compositional dimensionality, the possible number of ECIs within the interaction cutoffs is large. In addition, there are some constraints on the occupancies in the Li-Mn-Ti-O system, such as (1) the total number of Li, transition metals, and vacancies is fixed between octahedral and tetrahedral cation sublattice; (2) the net charge of the system must be neutral, etc. These relations and the inability to sample all possible configurations with DFT reduce the rank of the feature matrix below the dimension ($\text{rank}(\boldsymbol{\Pi}_S)=557$), which indicates that a sparse solution is required.

From the test results in Fig.\ref{fig:CV_L1vsL0L2}, we notice that when the sparseness varies, the variation of the CV error is smaller when the CE is regularized with the $\ell_0\ell_2$-norm than with the $\ell_1$-norm. This result indicates that $\ell_0\ell_2$ has a hyperparameter space that is larger and more tunable, whereas the $\ell_1$-CE is more deterministic with a small range of optimal $\mu_1$ obtained by minimizing the CV error. Motivated by this observation, the selection of ECIs for the Li--Mn--Ti--O system was completed as follows. 

The regularization strength $\mu_1$ in the $\ell_1$-CE was selected from the stable plateau region when minimizing the CV error in \textit{lasso} (e.g., the $\mu_1$ associated with points between sparseness of 50 to 100 in Fig.\ref{fig:CV_L1vsL0L2}). For the $\ell_0\ell_2$-norm, the $\mu_2$ was selected from the stable plateau region by minimizing the CV error in \textit{ridge regression}, similar to what is done for $\ell_1$-CE. After obtaining the optimal $\mu_2$, the solution for $\ell_0\ell_2$-CE was further determined by searching $\mu_0$ for a solution with the proper sparseness (at least $||\boldsymbol{J}||_0 < $ rank($\boldsymbol{\Pi}_S$), $\mu_1,\mu_2,\mu_0\in [10^{-6}, 10^{-1}]$). For both $\ell_1$-CE and $\ell_0\ell_2$-CE, several models with low CV error were tested for their ability to well reproduce physical properties, such as minimal violation of DFT ground states in the phase diagram, voltage profile comparison against DFT, as well as the Li-site energy difference between tetrahedral and octahedral occupancy. The best performing models for both $\ell_1$ and $\ell_0\ell_2$ are presented in Fig.\ref{fig:tina_CE}, respectively. 

Fig.\ref{fig:tina_CE}(a) presents a comparison of ground-state phase diagrams with the $\ell_1$-CE predictions, $\ell_0\ell_2$-CE predictions, and DFT calculations. The phase diagrams were generated with in-sample training data (all 1137 structures evaluated with DFT) for both DFT and CE models. We take the DFT phase diagram as the ground truth. In formation-energy prediction, the phase diagram is a key quantity that directly demonstrates the correct physics near the ground states. As the ground states are formed variationally, they are particularly discerning towards spurious ECIs, as the non-physical noisy interactions often create new ground states leading one to miss the true ground states. Thus, a well-reproduced phase diagram is desirable for a CE model. In our tests, the $\ell_1$-CE creates 12 new ground states, indicating that the correct physics in terms of cluster interactions is not well captured. However, the $\ell_0\ell_2$-CE preserves most of the DFT ground states, with only four spurious "ground states" in the $\ell_0\ell_2$-CE phase diagram. 

The improvement in the physics of the predictions associated with applying the $\ell_0\ell_2$-norm with hierarchy constraints is further demonstrated by the voltage profile and Li-occupancy energy. In Fig.\ref{fig:tina_CE}(b), the voltage profiles generated by prediction using the $\ell_1$-CE and $\ell_0\ell_2$-CE (blue lines) are compared with those from DFT (orange lines), taken as the ground truth. We can see that the $\ell_1$-CE incorrectly predicts the voltage plateau between $x=0.5$ to $1$ in the Li$_x$Mn$_2$O$_4$ spinel-like structure such that the $x=0.5$ configuration is no longer stable (the voltage between $x=0.5$ and $x=1.0$ is higher than that between $x=0.0$ and $x=0.5$). In contrast, the $\ell_0\ell_2$-CE matches very well with the DFT-generated voltage profiles. The erroneous predictions of the $\ell_1$-CE are further confirmed by the Li-occupancy energy. In Fig.\ref{fig:tina_CE}(c), the energy difference between Li in octahedral and tetrahedral occupancy was evaluated in the layered-MnO$_2$ and spinel-MnO$_2$ frameworks. The absolute error compared with DFT is 0.52 eV (layered) and 0.18 eV (spinel) for the $\ell_1$-CE, whereas that for the $\ell_0\ell_2$-CE is 0.09 eV (layered) and 0.09 eV (spinel), respectively. A significant reduction of prediction error is observed with the $\ell_0\ell_2$-norm regularized CE.

\section{Discussion \& Summary}
In two complex oxide systems, we showed that the $\ell_0\ell_2$-CE with hierarchy constraints outperforms the conventional $\ell_1$-CE in terms of sparseness against CV error, convergence rate with a subset of training-data, and some critical physical quantities in Li intercalation materials. More generally, the optimization of the ECIs is not deterministic within a single method, and the successful construction of a CE model typically relies on two aspects: (1) choosing a valid interaction space by truncating the clusters or orbits and (2) applying a proper optimization algorithm to obtain the ECIs. The results in this paper show that for the second step, the $\ell_0\ell_2$-norm method is the superior choice for a robust and physical solution compared to the conventional $\ell_1$ method.

We note that one limitation of the $\ell_0\ell_2$-norm method in the MIQP paradigm is the computational efficiency. As solving the $\ell_0$-norm is an NP-hard problem, more computational time is required to solve the MIQP when more predefined ECIs are included. The $\ell_0\ell_2$-CE works well for relatively small or well-predefined systems ($\mathrm{dim}(\boldsymbol{\Pi}_S) \leq 2000)$. Therefore, the most applicable way to use $\ell_0\ell_2$-norm regularized CE with hierarchy constraints is likely to be as follows: (1) define a CE within a relatively small cutoff and truncate to quadruplet or quintuplet clusters at most (ideally staying within $\mathrm{dim}(\boldsymbol{\Pi}_S) \leq 2000$) and (2) follow the procedure described in Fig.\ref{fig:structure_solver}(b) to determine the optimal hyperparameter to obtain the ECIs. However, we note that $\mathrm{dim}(\boldsymbol{\Pi}_S) \leq 2000$ applies to virtually all known published CE.

To obtain a model that represents the physics of a system well,  the relative difference of energies between configurations is of greater significance than the absolute ones. In ordinary least-squares fitting, the cost function only focuses on the global averaged error of the training set, which leads to over-fitting. Adding regularization of the ECIs can alleviate this issue by constraining the optimization space of parameters, but our results show that not all regularization creates physically meaningful solutions. We propose that it is beneficial to include the physically inspired constraints into the design of the cost function, such as adding hierarchy constraints with $\ell_0\ell_2$-norm implementation. The $\ell_0\ell_2$-CE can improve the physical meaning of the solution and break the correlation between coupled clusters, which is achieved by directly penalizing the number of non-zero ECIs for feature selection and enforcing hierarchy relations between ECIs via the slack variables in the MIQP paradigm. The $\ell_0\ell_2$-CE gives an estimation of the ECIs with reasonable physics near the ground-states, but does not strictly enforce the preservation of ground-states. Additionally, the ground-state preservation can be further achieved by adding inequality constraints on the energies into Eq.\eqref{eq:corrL0L2} as shown in previous work \cite{Huang2017GSpreserve}. 

In summary, our method sheds light on how to obtain good ECIs for simulations in complex and coupled multicomponent systems, with several proposed criteria in ECIs optimization: (1) minimize the CV error under general regression level (e.g., ridge regression); (2) as the sparseness describes the complexity of and number of independent variables, the sparseness of the solution shall be improved (reduced) with reasonable in-sample training error; and (3) check the near-ground-state behavior and related physics for the optimal ECIs selection.

\section{Acknowledgement}
This work was funded by the U.S. Department of Energy, Office of Science, Office of Basic Energy Sciences, Materials Sciences and Engineering Division under Contract No. DE-AC0205CH11231 (Materials Project program KC23MP). The work was also supported with computational resources provided by the Extreme Science and Engineering Discovery Environment (XSEDE), supported by National Science Foundation grant number ACI1053575; the National Energy Research Scientific Computing Center (NERSC), a U.S. Department of Energy Office of Science User Facility located at Lawrence Berkeley National Laboratory; the Center for Functional Nanomaterials (CFN), which is a U.S. Department of Energy Office of Science User Facility, at Brookhaven National Laboratory under Contract No. DE-SC0012704; and the Lawrencium computational cluster resource provided by the IT Division at the Lawrence Berkeley National Laboratory. 

\appendix*
\section{Appendix: DFT calculations}
DFT calculations were performed with the \textit{Vienna ab initio simulation package} (VASP) using the projector-augmented wave method \cite{kresse1996VASP, kresse1999PAW}, a plane-wave basis set with an energy cutoff of 520 eV, and a reciprocal space discretization of 25 \textit{k}-points per \AA. All the calculations were converged to $10^{-6}$ eV in total energy for electronic loops and 0.02 eV/Å in interatomic forces for ionic loops. In the LiF-MnO-LiVO$_2$-Li$_2$TiO$_3$ system, we used the Perdew-Burke-Ernzerhof (PBE) generalized gradient approximation exchange-correlation functional \cite{Perdew1996} with rotationally averaged Hubbard U correction (GGA+U) to compensate for the self-interaction error on all transition-metal atoms except titanium \cite{Wang2006}. The U parameters were obtained from the literature, where they were calibrated to transition-metal oxide formation energies (3.9 eV for Mn and 3.1 eV for V). The GGA+U computational framework is believed to be reliable in determining the formation enthapies of similar compounds \cite{Jain2011}. In the Li-Mn-Ti-O oxide system, the strongly constrained and appropriately normed (SCAN) meta-GGA exchange-correlation functional was used \cite{Sun2015SCAN}. The SCAN functional is believed to have better performance at capturing charge transfer due to better redox and atomic coordination prediction \cite{PRB_SCAN_Julia,zhang2018_npjSCAN}, which would improve the accuracy of energetics involving introducing vacancies on octahedral and interstitial tetrahedral sublattices in the rocksalt framework. 
\bibliography{solver.bib}

\end{document}